\documentclass[%
reprint,
 superscriptaddress,
nofootinbib,
amsmath,amssymb,
aps,
prb,
floatfix,
]{revtex4-2}

\usepackage{graphicx}
\usepackage{dcolumn}
\usepackage{bm}
\usepackage{xcolor}

\usepackage{mystyle}

\graphicspath{{figs/}} 

\begin{document}

\preprint{APS/123-QED}

\title{Superconductivity near two-dimensional Van Hove singularities: \\ a determinant quantum Monte Carlo study}


\author{Gustav Romare}
\affiliation{Department of Physics, University of Wisconsin-Madison, Madison, Wisconsin 53706, USA}
\author{Daniel Shaffer}
\affiliation{Department of Physics, University of Wisconsin-Madison, Madison, Wisconsin 53706, USA}
\author{Alex Levchenko}
\affiliation{Department of Physics, University of Wisconsin-Madison, Madison, Wisconsin 53706, USA}
\author{Edwin Huang}
\affiliation{Department of Physics and Astronomy, University of Notre Dame, Notre Dame, Indiana 46556, USA}
\author{Ilya Esterlis}
\affiliation{Department of Physics, University of Wisconsin-Madison, Madison, Wisconsin 53706, USA}

\date{\today}

\begin{abstract} 
The superconducting transition temperature $T_c$ of the two-dimensional attractive Hubbard model is computed in the vicinity of both ordinary (logarithmic) and higher-order (power-law) Van Hove singularities using determinant quantum Monte Carlo simulations. For interaction strengths $|U| \lesssim W/3$, where $W$ is the electronic bandwidth, $T_c$ is enhanced in the neighborhood of the Van Hove point, albeit more weakly than expected from weak-coupling BCS theory. Enhancing the Van Hove singularity from logarithmic to power-law yields only a minor additional enhancement of $T_c$. For $|U| \gtrsim W/3$, the maximum $T_c$ shifts away from the Van Hove point and instead occurs at a density unrelated to any features in the non-interacting density of states, consistent with a strong-coupling interpretation. We find that the maximal $T_c$ in the model is achieved at intermediate $U$ and at a density away from  the Van Hove point. 
\end{abstract}

\maketitle

\section{Introduction}

The appearance of Van Hove singularities (VHSs) \cite{vanhove1953} in the electronic density of states (DOS) can strongly influence the properties of low-dimensional metals. In two dimensions (2D), a saddle point in the electronic dispersion $\epsilon_\bfk$ generically leads to a logarithmic divergence in the DOS. Further flattening of the dispersion produces a ``higher-order" VHS (HOVHS, also referred to as an extended VHS in earlier literature), characterized by even stronger power-law divergences that have been observed in some experiments \cite{Abrikosov93, shtyk2017, efremov2019, yuan2019, isobe2019, yuan2020, chandrasekharan2020, classen2020, HuThomale22, ojajarvi2024, chandrasekaran2024, ClassenBetouras24, PullasseriSantos24, WangSantos25}. Interaction effects are expected to become enhanced when the Fermi energy lies near such VHSs (both ordinary and higher-order), increasing the propensity toward ordering. Indeed, the potential to engineer higher-$T_c$ superconductors by tuning the Fermi level to a VHS has attracted sustained interest over several decades \cite{hirsch1986, radtke1993, radtke1994, Markiewicz97, ma2014, ojajarvi2024, HohmannThomale25}.

For weak electron-electron attraction, BCS calculations predict that the divergent DOS at a VHS leads to a strong enhancement of the superconducting $T_c$. At an ordinary VHS, the usual BCS transition temperature $T_c \sim \Lambda \exp(-1/\lambda)$ is enhanced to $T_c \sim \Lambda\exp(-1/\sqrt\lambda)$, where $\lambda$ is an appropriate dimensionless coupling strength and $\Lambda$ is a high-energy cutoff on the interaction \cite{hirsch1986}. At a HOVHS with DOS divergence $\sim |\epsilon|^{-1/4}$, the BCS transition temperature goes as a power law, $T_c \sim \lambda^4$.  

While there is little doubt that proximity to VHSs is an effective way of increasing the $T_c$ of weakly-coupled, ``low-$T_c$" superconductors, an important question is the extent to which DOS singularities remain similarly important away from weak coupling, where finite quasiparticle lifetimes smear the VHS and the pairing vertex is renormalized by interactions. Of course, for sufficiently strong interactions, of the order of the electronic bandwidth, strong-coupling effects will wash out any details of the Fermi surface entirely. 
This question is especially important in analyzing the role of VHSs for superconductivity in many experimental systems of current interest, such as strontium ruthenate Sr$_2$RuO$_4$ \cite{steppke2017,sunko2019, Mueller24}, CeRh$_2$As$_2$ \cite{LeeAgterberg25}, ${\mathrm{Ti}}_{4}{\mathrm{Ir}}_{2}\mathrm{O}$ \cite{WuAgterberg25},  kagome superconductors \cite{HuThomale22, Kang2022, WuRaghu23, Luo23}, perovskites \cite{Li24}, 3D nickelates \cite{Xia25}, and van der Waals materials \cite{Li2010,Xie2019,Kerelsky2019,Choi2019,Jiang2019,WanUgeda23}. More generally, the effectiveness of VHSs is important for understanding what system parameters should be optimized to achieve higher-$T_c$ 2D superconductivity, which inevitably requires going away from weak coupling.

There is a significant body of theoretical literature addressing the question of superconductivity near VHSs, via both numerical \cite{hirsch1986,radtke1993, radtke1994, ma2014} and analytical methods. Most of the attention, initially motivated by the discovery of high-$T_c$ superconductors, has focused on the repulsive Hubbard model studied using the parquet renormalization group \cite{Dzyaloshinskii87, Markiewicz97, FurukawaSalmhofer98, HonerkampSalmhofer01, NandkishoreLevitovChubukov12,  MaitiChubukov13}. Similar scenarios with emergent effective attractive pairing interactions in the presence of VHSs have been explored in graphene \cite{NandkishoreLevitovChubukov12} and moir\'{e} graphene systems \cite{ClassenHonerkampScherer19, ChichinadzeClassenChubukov20, HsuDasSarma20, WuWuWu23}, as well as in systems with topologically non-trivial Chern bands \cite{ShafferSantos22, ShafferSantos23}, rhombohedral graphene \cite{ChouDasSarma25}, and altermagnets \cite{RaoClassen25, PartheniosClassen25}.
More recently, the possibility of HOVHSs in the DOS as a way to achieve yet higher transition temperatures has also been actively investigated theoretically
\cite{LinNandkishore20, AksoyChamon23, ojajarvi2024, ClassenBetouras24}, including in moir\'{e} transition metal dichalcogenides \cite{HsuDasSarma21}, the Haldane model \cite{CastroShafferWuSantos23}, Bernal bilayer graphene \cite{LeeChichinadzeChubukov24}, and cuprates \cite{MarkiewiczBansil25}.  Although early works noted the reduced effectiveness of the ordinary VHS \cite{hirsch1986, radtke1993} and HOVHS \cite{radtke1994} mechanisms for enhancing $T_c$ with increasing  coupling, the relevance of the VHS scenario beyond weak-coupling remains largely unexplored, even in models with explicitly attractive interactions.

 In the present paper, we address the question of superconductivity near both ordinary and higher-order VHSs away from weak coupling, using numerically exact determinant quantum Monte Carlo (DQMC) simulations of the 2D attractive Hubbard model. While the 2D attractive Hubbard model is not a realistic model for any real 2D or quasi-2D material, it has the virtue that, generically, there are no competing instabilities to $s$-wave superconductivity. We are thus able to isolate the effect of VHSs on superconductivity specifically, and systematically track the evolution of $T_c$ from weak to strong coupling. Moreover, the model is free from the notorious fermion minus-sign problem \cite{loh1990} for any electron density and thus can be simulated efficiently by DQMC for large system sizes and down to low temperatures, over a range of densities around the VHS.
 
 As we demonstrate below, even when the interaction strength is relatively weak (a fraction of the electronic bandwidth), the effect of VHSs in enhancing the superconducting $T_c$ is significantly reduced compared to expectations based on weak-coupling (unrenormalized) BCS theory. Nevertheless, as long as the coupling is not too large, we find that $T_c$ as a function of electron density exhibits a well-defined maximum near the VHS density, and the evolution of $T_c$ can be captured by direct perturbation theory in the interaction strength when both self-energy effects and corrections to the pairing vertex are taken into account. Beyond a critical coupling strength of order $|U| \sim W/3$, where $W$ is the electronic bandwidth, the effects of the VHS are rapidly washed out, and the position of the $T_c$ maximum moves away from the VHS density with increasing $|U|$. Of course, for values of $|U| \gg W$, this evolution is to be expected, as the system crosses over into the regime of preformed local pairs and Bose-Einstein condensation (BEC) \cite{nozieres1985, micnas1990, melo2024}, where the transition temperature decreases as $T_c \sim t^2/|U|$ (we recall that, in the $s$-wave case, the evolution from BCS to BEC pairing is continuous). What is surprising, however, is how sudden the crossover is from ``Fermi-surface pairing" to ``local pairing". For the model parameters used (see below), we find that the maximal $T_c$ as a function of $U$ and density $n$ occurs at the intermediate value $U \approx -6 t$ and at a density far from the VHS, not tied to any features of the non-interacting band structure.  

\section{Model and methods} 

\subsection{Hamiltonian}

\begin{figure}
    \centering
    \includegraphics[width=\linewidth]{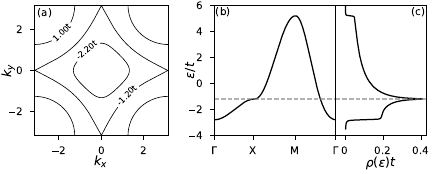}
    \includegraphics[width=\linewidth]{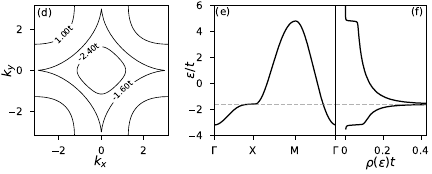}
    \caption{Band structure used in this study. Upper figures are for $t'=-0.3t$ and $t''=0$ (ordinary VHS) and bottom figures are for $t'=-0.3t$ and $t''=0.1t$ (HOVHS). (a,d) Fermi contours near the VHSs. (b,e) Dispersion along a triangular path in the Brillouin zone. (c,f) Density of states showing the position of the VHSs. Note the asymmetry of the divergence for the HOVHS.}
    \label{fig:fermiology}
\end{figure}

We have studied the attractive Hubbard model
    \be
    H = \sum_{\bfk\sigma} \epsilon_\bfk c^\dag_{\bfk\sigma} c_{\bfk\sigma} + U \sum_i n_{i\uparrow}n_{i\downarrow},
    \label{eq:ham}
    \ee
where $c^\dag_{\bfk\sigma}$ creates an electron with lattice momentum $\bfk$ and spin $\sigma$, $n_{i\sigma} = c^\dag_{i\sigma}c_{i\sigma}$ is the local electron density operator on site $i$, and the interaction $U = -|U| < 0$.  We work on a 2D square lattice, and the most general form of electronic dispersion  we use includes both first, second, and third neighbor hoppings, $t$, $t'$, and $t''$, respectively:
\be
\begin{aligned}
    \epsilon_\bfk &= -2t (\cos k_x + \cos k_y) 
    -4t' \cos k_x \cos k_y \\ 
    &\qquad -2t'' (\cos 2k_x + \cos 2k_y).
\end{aligned}
     \label{eq:disp}
\ee
For $t''=0$, the second neighbor hopping $t'$ is needed to move the (logarithmic) VHS away from perfect nesting of the Fermi surface, where the degeneracy between superconductivity and charge-density wave (CDW) order leads to a three-component order parameter and a vanishing $T_c$ by the Mermin-Wagner theorem\footnote{We have verified that the CDW correlations remain short-ranged for all parameters considered in this study; see Appendix~\ref{app:CDW}.}\cite{hirsch1985}. The third-neighbor hopping $t''$ is included to tune the coefficient of $k_x^2$ in the expansion of \eqref{eq:disp} about the saddle-point to zero by setting $t'' = (t+2t')/4$, thus yielding a HOVHS with DOS divergence $\sim |\epsilon|^{-1/4}$. Note that, for the HOVHS, the DOS divergence is asymmetric, with different coefficients depending on whether the singularity is approached from above or below. For concreteness, we will specialize to $t' =-0.3t$, $t''=0$ for the ordinary VHS, and $t' = -0.3t$, $t''=0.1t$ for the HOVHS. The non-interacting bandwidth is $W = 8t$ in both cases. The corresponding band structures and DOS are summarized in Fig.~\ref{fig:fermiology}. We analyze interaction strengths $ |U| \leq 8 t$.

\subsection{Numerical Methods}
We have utilized DQMC simulation \cite{blankenbecler1981,hirsch1985,white1989,scalettar1989,moreoscalapino_scaling} to study the model \eqref{eq:ham}. We have studied square lattices with $N=L^2$ sites, up to linear system size $L = 22$ with periodic boundary conditions, down to temperatures $T/t = 1/\beta t \approx 0.03$. Finite-size effects are reduced by threading a small magnetic flux through the system \cite{assaad2002}. For more details on the DQMC simulation see Appendix \ref{app: finite-size-effects-high-U}. 

Quasi-long-range superconducting order sets in below the Berezinskii-Kosterlitz-Thouless (BKT) temperature $T_c$, which can be extracted from the superfluid density $\rho_s$ via the BKT criterion \cite{nelson1977,paiva2004}
\begin{equation}
    \rho_s(T_c^{-}) = \frac{2}{\pi}T_c.
    \label{eq:BKT}
\end{equation}
Here $\rho_s(T_c^-)$ is the value of the superfluid density as $T$ approaches $T_c$ from below. The superfluid density can be obtained directly from DQMC data according to \cite{scalapino1993}
\begin{equation}
    \rho_s = \frac{1}{4} \left[\langle K_{xx} \rangle - \Lambda_{xx}(\bfq\to0, i\nu_m=0) \right],
    \label{eq:rho_s}
\end{equation}
where $\langle K_{xx} \rangle$ is the expectation value of the diamagnetic term and $\Lambda_{xx}(\bfq, i\nu_m)$ is the current-current correlation function 
    \be
    \Lambda_{xx}(\bfq, i\nu_m) = \frac 1N \int_0^\beta \dd\tau ~ e^{i\nu_m \tau} \langle j_x^P(\bfq, \tau) j_x^P(-\bfq, 0)  \rangle,
    \ee
with bosonic Matsubara frequency $\nu_m = 2\pi m T$. The diamagnetic term $K_{xx}$ and the $x$-component of the paramagnetic current operator are
    \begin{align}
    K_{xx} &= -\frac{1}{N} \sum_{l \delta \sigma} t_{l+\delta, l}\delta_x^2(c^{\dagger}_{l+\delta,\sigma}c^{}_{l\sigma} + \text{h.c.}), \\
    j^P_x (\bfq) &= \sum_{\bfr} e^{-i\bfq \cdot \bfr} j^P_x(\mathbf r), \\
    j^P_x(\mathbf r_l) & = i\sum_{\delta \sigma} t_{l+\delta, l}\delta_x(c^{\dagger}_{l+\delta,\sigma}c^{}_{l\sigma} - \text{h.c.}),
    \end{align}
where $\delta \in \{(1,0),\,(1,1),\,(1,-1),\,(2,0)\}$, and $t_{ij}$ is the hopping matrix with non-zero elements $t,\,t',\,t''$.
By numerically solving Eq.~\ref{eq:BKT} we get finite-size estimates $T_c(L)$ that can be used to extrapolate to the thermodynamic limit $L \to \infty$ \footnote{A more rigorous treatment of the size effects involves known logarithmic finite-size corrections to $\rho_s$ for the BKT transition \cite{hsieh2013}. However, we find a linear extrapolation produces indistinguishable results from the more rigorous scaling form \cite{hsieh2013} and we have thus chosen to use the simpler linear extrapolation.}. The limit $\mathbf q \to 0, i\nu_m=0$ in Eq.~\ref{eq:rho_s} should be the transverse limit $(q_x = 0, q_y \to 0)$ \cite{scalapino1993}, but for the torus one can safely set $\mathbf q=0, i\nu_m=0$, as we explain in Appendix~\ref{app: Static Uniform limit}. The estimate for $T_c$ obtained from the BKT criterion in this way is in good agreement with $T_c$ obtained from finite size scaling of the equal time pair correlation function. 

The pair correlation function and additional data regarding finite-size effects are reported in Appendix~\ref{app:extrap}.

To get an estimate for the electronic DOS in the interacting system we use the imaginary-time proxy \cite{trivedi1995}
    \be
    \rho_\beta \equiv -\frac \beta \pi G(\mathbf r = 0, 
    \tau = \beta/2) = \frac{\beta}{2\pi}\int d\omega \frac{\rho(\omega)}{\cosh(\beta\omega/2)},
    \label{eq:DOS_proxy}
    \ee
where $\rho(\omega) = -\im G_R(\bfr = 0, \omega)/\pi$ is the local DOS ($G_R$ is the retarded electron Green's function) and $G(\mathbf r, \tau > 0) = -\langle c(\mathbf r,\tau)c^\dagger(0, 0)\rangle$ is the imaginary-time, real-space Green's function directly sampled in the DQMC simulation. The proxy $\rho_\beta$ is essentially an average of the low-energy DOS over an energy window of order $\sim T$, and is expected to yield a reliable estimate as long as the DOS does not possess strong features on energy scales less than $T$. As an alternative DOS proxy, we have also measured the uniform static spin susceptibility. 

The spin susceptibility agrees qualitatively with $\rho_\beta$, as we verify in Appendix ~\ref{app:DOS_proxies}. 

\begin{figure}
    \centering
    \large $\Sigma$ = \includegraphics[width=0.1\linewidth]{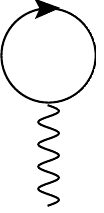} + 
    \includegraphics[width=0.2\linewidth]{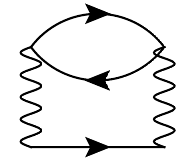}\\
    ~\\
    \includegraphics[width=0.75\linewidth]{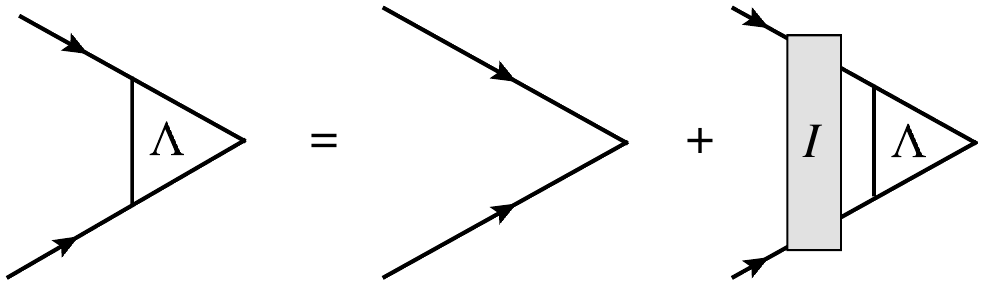} \\
    ~\\
    \includegraphics[width=0.75\linewidth]{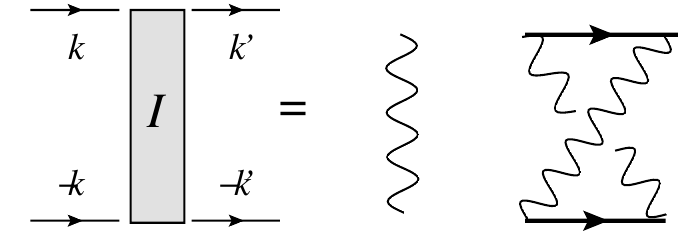} 
    \caption{Diagrams contributing to the electronic self-energy $\Sigma$ and irreducible pairing vertex $I$ through second order in the interaction $U$.}
    \label{fig:diagrams}
\end{figure}

\subsection{Perturbation theory} 
\label{method:pert_theory}
To help interpret the DQMC results in the regime of weak to intermediate coupling, we have performed perturbative calculations of the electronic self-energy, pair susceptibility, and superconducting $T_c$. The standard dimensionless coupling governing the perturbative expansion is  $U\rho_0$,  where $\rho_0$ is the DOS at the Fermi level. Although this coupling is not small near a VHS, the dimensionless parameter $\lambda = |U|/W$ is still a useful measure of the interaction strength at non-zero temperature. Our studied range of $|U| \leq 8t$ corresponds to $\lambda \leq 1$; as described above, when $|U| \gtrsim W$, the system crosses over into the BEC regime of preformed pairs and the perturbative approximation fails even qualitatively.

We consider the self-energy diagrams shown in Fig.~\ref{fig:diagrams}, which give the self-energy through second order in $U$ as
    \be
    \Sigma(k) = \frac 12 U n -\frac{U^2 T}{N} \sum_{k'} \chi_0(k-k')G_0(k').
    \label{eq:sig2}
    \ee
We adopt the shorthand $k = (\bfk, i\omega_m)$ with fermionic Matsubara frequencies $\omega_m = (2m+1)\pi T$ and the sum on $k'$ denotes a summation over wavevectors $\bfk'$ and internal Matsubara frequencies $\omega_{m'}$. The function $\chi_0$ is the non-interacting particle-hole susceptibility
    \be
    \chi_0(q) = \frac{1}{\beta N}\sum_{k} G_0(k)G_0(k+q),
    \ee
where $q = (\bfq, i\nu_m)$. 

The perturbative self-energy is used in the equation for the pairing vertex $\Lambda$, which is expressed in terms of the irreducible vertex $I$ as
    \be
    \Lambda(k) = 1  - \frac{1}{\beta N}\sum_{k'} I(k,k') G(k')G(-k')\Lambda(k').
    \label{eq:lam}
    \ee
To second-order in $U$, the electron Green's function $G(k)$ is dressed with the second-order self-energy \eqref{eq:sig2}, and the irreducible vertex through order $U^2$ is
    \be
    I(k,k') = U - U^2 \chi_0(k-k').
    \label{eq:I}
    \ee
The corresponding diagrams are shown in Fig.~\ref{fig:diagrams}.
In terms of the pairing vertex, the pair susceptibility is given by
    \be
    P_s = \frac{1}{\beta N}\sum_k  G(k) G(-k) \Lambda(k),
    \ee
and superconducting $T_c$ is determined by the divergence of $P_s$. When $T \to T_c^+$, Eq.~\eqref{eq:lam} becomes the eigenvalue equation
    \be 
    -\frac{1}{\beta N}\sum_{k'} I(k,k') G(k')G(-k')\Lambda(k') = \lambda(T) \Lambda(k).
    \label{eq:gap_eq}
    \ee
Within this approximation, the superconducting transition occurs when the maximal eigenvalue $\lambda_\text{max}$ of the above equation goes to unity, $\lambda_\text{max}(T_c) = 1$. Note that the standard, unrenormalized BCS approximation is obtained from Eq.~\eqref{eq:gap_eq} by setting $G = G_0$ and $I(k,k') = U$. The random-phase approximation (RPA) is obtained by retaining self-energy corrections but neglecting the vertex corrections (i.e., still setting $I(k,k') = U$), corresponding to summing the ladder series for the pair susceptibility with dressed electronic Green's functions. 

We will demonstrate below that, for $\lambda \lesssim 1/4$, the perturbative treatment including both self-energy and vertex corrections is in satisfactory agreement with the DQMC calculations. As noted in \cite{hirsch1986}, both the self-energy and the vertex corrections are detrimental to $T_c$: the former leads to the smearing of the DOS and broadens the peak in $T_c$ as a function of $n$, while the latter decreases the effective interaction strength (note that $\chi_0<0$, such that $|I|<|U|$ in Eq. \ref{eq:I}). Since the VHS also increases the magnitude of $\chi_0$ (without leading to a divergence in the regime we consider, i.e., in the absence of perfect nesting), the VHS actually exacerbates these effects. This may qualitatively explain why the $T_c$ enhancement becomes severely limited at already moderate interactions strengths, as we report in the next section.

\section{Results} 

\begin{figure}[h!]
    \centering
    \includegraphics[width=\linewidth]{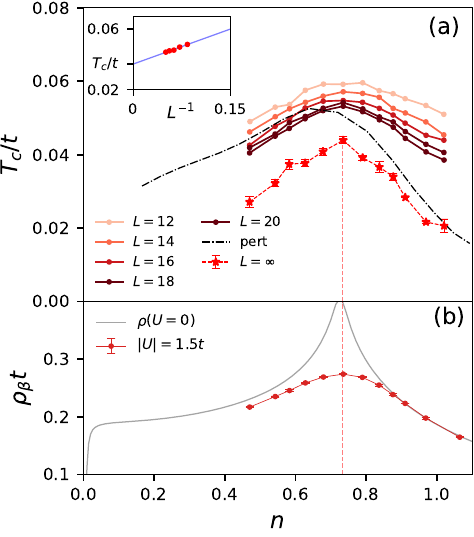}
    \caption{(a) Superconducting $T_c$ near an ordinary VHS at $U/t=-1.5$ for finite system sizes, together with $T_c$ extrapolated to the thermodynamic limit $L\to\infty$. The inset shows the extrapolation procedure, using the VHS density as an example. Also shown is $T_c$ obtained in second-order perturbation theory. (b) DOS proxy $\rho_\beta$ as a function of density. The solid line is the non-interacting DOS.} 
    \label{fig:Tc_oVH}
\end{figure}

\begin{figure}[h!]
    \centering
    \includegraphics[width=\linewidth]{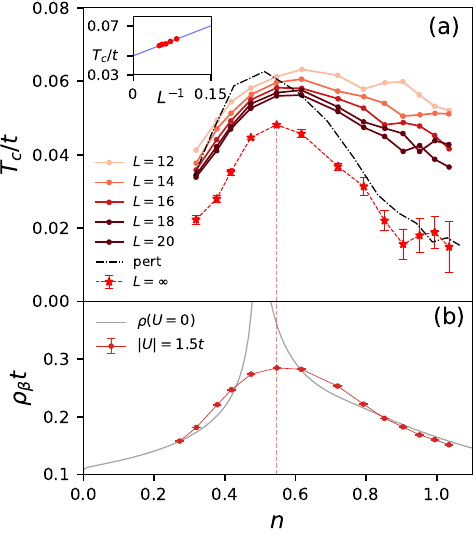}
    \caption{(a) Superconducting $T_c$ near a HOVHS for $U/t=-1.5$ for finite system sizes, together with $T_c$ extrapolated to the thermodynamic limit $L\to\infty$. The inset show the extrapolation procedure, using the HOVHS density as an example. Also shown is $T_c$ obtained in second-order perturbation theory. (b) DOS proxy $\rho_\beta$ as a function of density. The solid line is the non-interacting DOS.} 
    \label{fig:Tc_hVH}
\end{figure}

We now present our results, organized according to interaction strength. We begin in the weak-to-intermediate coupling regime, where the effects of VHSs are expected to be most pronounced, and then track the evolution toward intermediate and strong coupling. 
In Figs.~\ref{fig:Tc_oVH} and \ref{fig:Tc_hVH} we show the superconducting $T_c$ as a function of density around the ordinary VHS and HOVHS, respectively, for interaction strength $U=-1.5t$. Insets of the figures show the extrapolation of $T_c$ to the thermodynamic limit $L\to\infty$. Data for the superfluid stiffness used to obtain $T_c(L)$ is provided in Appendix~\ref{app:extrap}. Also shown in Figs.~\ref{fig:Tc_oVH}b and \ref{fig:Tc_hVH}b is the DOS proxy $\rho_\beta$ (Eq.~\ref{eq:DOS_proxy}) as a function of density, which we use to identify the location of the VHS in the interacting system. Data for $U=-2t$ are similar and are reported in  Appendix~\ref{app:analytics} (see also Fig.~\ref{fig:Tcs_vs_n_higher_Us}).

For $|U|\lesssim 2t$, we find that $T_c$ is maximal near the VHS density for both the ordinary and higher-order VHS, with the maximum becoming broader with increasing strength of the interaction (this can be seen most clearly in Fig.~\ref{fig:Tcs_vs_n_higher_Us}b). Note that the asymmetry of the HOVHS divergence leads to a shift in the $T_c$ maximum with varying $U$; however, the $T_c$ maximum continues to track the DOS proxy. For $|U|=1.5t$, comparison of Figs.~\ref{fig:Tc_oVH} and \ref{fig:Tc_hVH} shows that the enhancement of $T_c$ upon going from the ordinary to higher-order VHS is relatively weak and the difference diminishes further as the interaction strength increases. 

The DQMC $T_c$ is also compared with estimates from second-order perturbation theory (as described in Sec.~\ref{method:pert_theory}) in Figs.~\ref{fig:Tc_oVH} and \ref{fig:Tc_hVH}. The agreement is satisfactory over much of the density range shown. We have found that this level of agreement requires the inclusion self-energy and vertex corrections, both of which tend to reduce $T_c$. In Appendix~\ref{app:analytics} we also compare the measured $T_c$ with BCS and RPA predictions, demonstrating how self-energy and vertex corrections contribute to both reducing the absolute value of $T_c$ and broadening the maximum around the VHS density. 

\begin{figure}
    \centering
    \includegraphics[width=\linewidth]{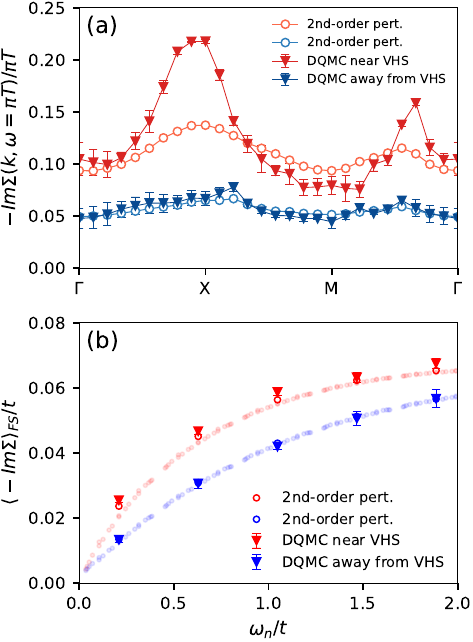}
    \caption{(a) Imaginary part of the electronic self-energy at lowest Matsubara frequency $\omega_m = \pi T$, normalized by $\pi T$, as a function of $\bfk$ through a path in the Brillouin zone. (b) Imaginary part of the self-energy averaged over the Fermi surface as a function of Matsubara frequency. In panels (a) and (b) we show data for $n\approx 0.72$ (at the ordinary VHS) and $n\approx 1$ (away from the VHS). Here $|U| = 1.5t$ and $\beta t =15$. In both panels, DQMC results are compared with second-order perturbation theory. In panel (b), we have included the results of perturbation down to lower temperatures $15\le \beta t \le 90$ to demonstrate that the self-energy has essentially converged to its $T\to 0$ limit. }
    \label{fig:sig}
\end{figure}

\begin{figure}
    \centering
    \includegraphics[width=\linewidth]{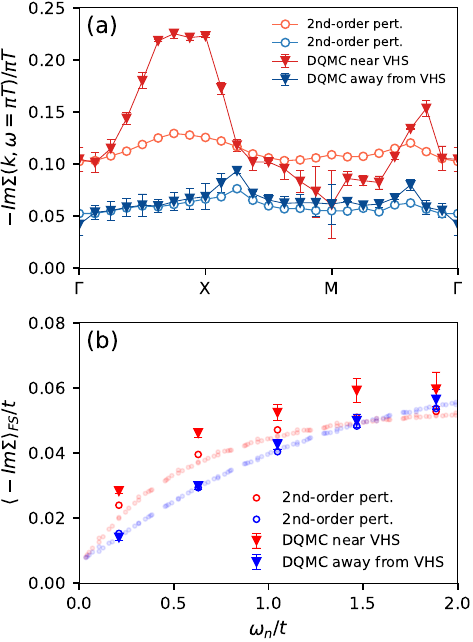}
    \caption{(a) Imaginary part of the electronic self-energy at lowest Matsubara frequency $\omega_m = \pi T$, normalized by $\pi T$, as a function of $\bfk$ through a path in the Brillouin zone. (b) Imaginary part of the self-energy averaged over the Fermi surface as a function of Matsubara frequency at the HOVHS. In panels (a) and (b) we show data for $n\approx 0.55$ (at the HOVHS) and $n\approx 1.0$ (away from the HOVHS). Here $|U| = 1.5t$ and $\beta t =15$. In both panels, DQMC results are compared with second-order perturbation theory. In panel (b), we have included the results of perturbation down to lower temperatures $15\le \beta t \le 90$ to demonstrate that the self-energy has essentially converged to its $T\to 0$ limit.}
    \label{fig:sig_hvh}
\end{figure}

We have also extracted the electronic self-energy from our DQMC simulations, and compared it with that obtained from perturbation theory (Eq.~\eqref{eq:sig2}). The imaginary part of the self-energy is shown Figs.~\ref{fig:sig} and \ref{fig:sig_hvh}, corresponding to the ordinary and higher-order VHS, respectively. Figs.~\ref{fig:sig}a  and \ref{fig:sig_hvh}a show the self-energy at the lowest Matsubara frequency ($\omega_m = \pi T$) as a function of $\bfk$ for densities both at, and away from, the VHS. As expected, the self-energy is noticeably larger at the VHS density. The perturbative approximation underestimates the magnitude of the self-energy near the VHS in the Brillouin zone (the X point), but gives a good quantitative approximation away from that point. For densities away from the VHS, perturbation theory accurately captures the self-energy over the entire range of $\bfk$. These observations apply to both the ordinary VHS and HOHVS. 

In Figs.~\ref{fig:sig}b and \ref{fig:sig_hvh}b, we report the Fermi surface averaged self-energy as a function of Matsubara frequency. The averaged quantities agree well between perturbation theory and DQMC, suggesting that, despite the underestimate of the self-energy at the VHSs in the Brillouin zone, quantities that depend on averages over the entire Fermi surface (relevant for $s$-wave superconductivity) are likely to be captured reasonably well by perturbation theory. It will be interesting to investigate how the $\bfk$-space structure is reflected in situations where the order parameter has non-trivial momentum dependence, e.g., for $d$-wave pairing. Additional data for the self-energy at stronger coupling are reported in Appendix~\ref{app: strong coupling self-energy}. 

\begin{figure}[h!]
    \centering
    \includegraphics[width=\linewidth]{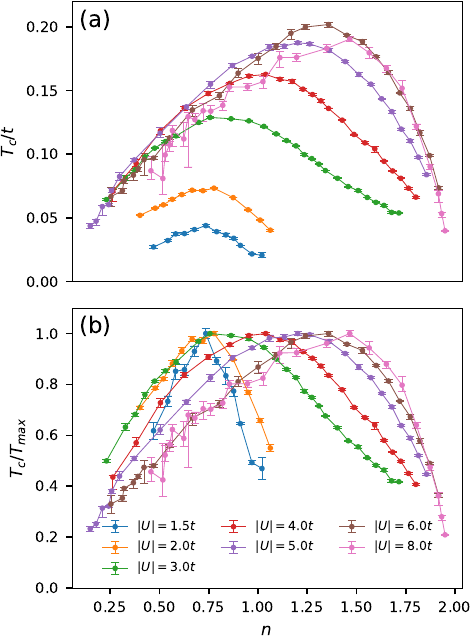}
    \caption{(a) Superconducting $T_c$ as a function of density $n$ for different values of $U$. (b) Same data but normalized by the maximal value $T_{c,\text{max}}$ for each $U$. Here $t''=0$, corresponding to an ordinary VHS in the non-interacting DOS. For $|U|\lesssim 4t$, there is a maximum in $T_c$ near the VHS fillings, which is broadened with increasing $|U|$. For $|U| \gtrsim 4t$,  the maximal $T_c$ shifts rapidly to a density $n \approx 1.35$, which is away from the VHS and appears unrelated to any Fermi surface features.}
    \label{fig:Tcs_vs_n_higher_Us}
\end{figure}

We now describe the evolution of $T_c$ for stronger coupling. Given the qualitative similarity between the ordinary and HOVHS cases, we restrict our attention here to the ordinary VHS (corresponding to $t''=0$). We have also found that, for $|U| > 2t$, finite-size effects on $T_c$ are much weaker than at smaller coupling. Correspondingly, the data shown for $|U| =3t$ and $|U|=4t$ are for $L=16$, while for larger $|U|$ we use $L=12$. These system sizes are large enough that size effects appear negligible; further details on the size-dependence are given in Appendix~\ref{app: finite-size-effects-high-U}. 
Our results for $T_c$ in the range $1.5t \leq U \leq 8t$ are reported in Fig.~\ref{fig:Tcs_vs_n_higher_Us}. For $|U| > 2t$,  the $T_c$ maximum broadens significantly and shifts abruptly away from the VHS\footnote{At these larger values of $U$, $\rho_\beta$ becomes strongly temperature dependent and ceases to be a useful proxy for the interacting DOS; see Appendix \ref{app:DOS_proxies}.}. To highlight the evolution of the width of the $T_c$ maximum, we have normalized $T_c$ by its maximal value at each $U$ in Fig.~\ref{fig:Tcs_vs_n_higher_Us}b. These results indicate that the breakdown of VHS-enhanced superconductivity occurs already at intermediate coupling, well before the system enters the extreme strong-coupling regime. The rapid crossover may reflect a breakdown of perturbation theory due to singularities in the irreducible vertex function, as seen in dynamical mean-field theory studies of both attractive \cite{crossover1} and repulsive \cite{crossover2} Hubbard models.

\begin{figure}
    \centering
    \includegraphics[width=\linewidth]{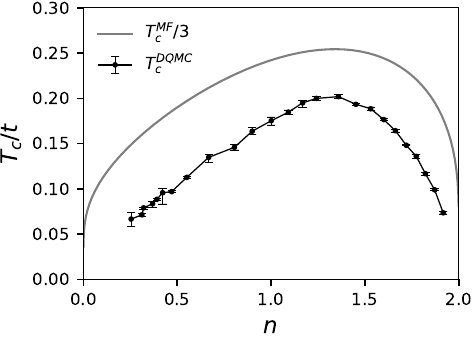}
    \caption{Comparison of $T_c$ obtained by DQMC and the strong-coupling, mean-field prediction Eq.~\ref{eq:Tc_MF}. The mean-field curve has been rescaled by a factor of $1/3$ for visual clarity. Here $|U|=6t$.}
    \label{fig:strong_coupling_MF_VS_DQMC}
\end{figure}

For $|U| \approx 8t$, we find that the $T_c$ maximum remains near the density $n_* \approx 1.35$. This new maximum can be rationalized from a strong-coupling expansion in powers of $t^2/U$, in which the system may be described in terms of an effective hard-core Bose Hubbard (or equivalently pseudospin-1/2) model \cite{micnas1990,Freericks1998}. We have carried out a mean-field analysis of the resulting effective Hamiltonian, with the mean-field $T_c$ given by (see Appendix~\ref{app: Strong coupling expansion} for details)
    \be
    T_c^\text{MF} = \frac{(n-1)}{\ln\left(\frac{n}{2-n}\right)}\frac{8t^2}{|U|}\left[1+\left( \frac{t'}{t}\right)^2 +\frac{6t'}{|U|}(1-n)\right].
    \label{eq:Tc_MF}
    \ee
The third order term, proportional to $t^2t'/|U|$, introduces an asymmetry in $T_c$ about half-filling $n=1$. The transition temperature $T_c^\text{MF}$  is compared with the DQMC results in Fig.~\ref{fig:strong_coupling_MF_VS_DQMC}. The position of the $T_c$ maximum and the overall shape of the $T_c(n)$ curve agree reasonably well with the strong-coupling prediction. As is to be expected, the mean-field approximation overestimates the overall magnitude of transition temperature. We note that Eq.~\eqref{eq:Tc_MF} implies that the $T_c$ maximum should eventually move toward $n=1$ for sufficiently large $|U|$. With increasing $|U|$, we therefore expect a non-monotonic evolution of the density at which $T_c$ is maximal,  shifting from  $n_\text{VH}\approx 0.72$ to $n_* \approx 1.35$, and eventually to  $n=1$. 

\begin{figure}[h!]
    \centering
    \includegraphics[width=\linewidth]{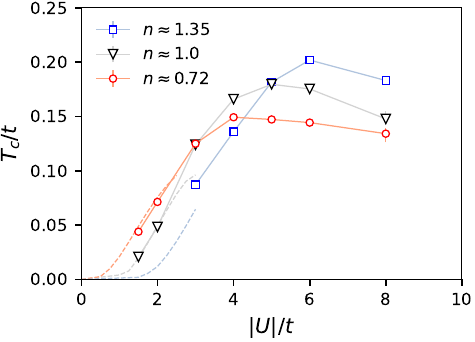}
    \caption{Superconducting $T_c$ as a function of $|U|$ at the VHS density $n\approx 0.72$ (red), optimal density $n\approx 1.35$ (blue), and an intermediate density $n\approx1.0$ (black). The dashed lines correspond to $T_c$ calculated from second-order perturbation theory. Here $t''=0$, corresponding to an ordinary VHS in the non-interacting DOS. We find a global maximum in $T_c$ at $n \approx 1.35$ and $|U| \approx 6t$.}
    \label{fig:Tcs_vs_U}
\end{figure}

Finally, we report the superconducting $T_c$ as a function of $|U|$ in Fig.~\ref{fig:Tcs_vs_U}, for several representative densities. Here, we exploit the good agreement between $T_c$ computed via perturbation theory and DQMC to estimate $T_c$ at weaker coupling---where perturbation theory is expected to be more accurate---and where we cannot go to low enough temperatures in the DQMC simulations to reach the superconducting state\footnote{For $|U| = t$, we have verified that the pair susceptibility $P_s$ computed with DQMC agrees with that obtained from the perturbation theory down to the lowest accessible temperatures.}. From Fig.~\ref{fig:Tcs_vs_U} it is clear that, for sufficiently weak coupling, proximity to the VHS density yields the optimal $T_c$. With  increasing interaction strength, however,  the influence of the VHS is rapidly suppressed, and $T_c$ is maximized at densities away from the VHS. For the choice of hopping $t'=-0.3t$ discussed here, we find a global maximum in $T_c$ for the Hamiltonian~\ref{eq:ham} at $n \approx 1.35$ and $U \approx -6t$ (see also Appendix~\ref{app: Strong coupling expansion}).

\section{Conclusions}

For interaction strengths up to $ |U| \lesssim W/3$, we verify that proximity to a VHS enhances $T_c$, albeit more weakly than expected from weak-coupling BCS theory. For the interaction strengths considered, we find only a modest enhancement of $T_c$ when the VHS is strengthened from logarithmic to power-law. Comparison with perturbation theory shows that both the reduction in $T_c$ and the broadening of the its maximum near the VHS can be accounted for by self-energy and vertex corrections. 

For $|U| \gtrsim W/3$, the influence of the VHS is rapidly suppressed. While this is expected at sufficiently large $|U|$, where the system crosses over from the BCS to BEC pairing, we find that the crossover is abrupt, with a strong-coupling expansion already providing a qualitatively accurate description for $|U| \approx W$. The maximal $T_c$ occurs at intermediate coupling $U \approx -6t$ and a density away from the VHS, consistent with strong-coupling theory. While $T_c$ is generically expected to be maximal at intermediate $U$, the precise value of the optimal density depends sensitively on the hopping parameters; this point is further discussed in Appendix~\ref{app: Strong coupling expansion}.

The results presented here are obtained within the attractive Hubbard model, which is at best a crude approximation to real materials. Nevertheless, the limited effectiveness of VHSs in enhancing $T_c$ at both strong and intermediate coupling is likely a general phenomenon. Similar conclusions have been reached within the Migdal-Eliashberg framework for electron-phonon systems\footnote{Within the Migdal-Eliashberg theory, vertex corrections are neglected by Migdal's theorem, and the suppression of $T_c$ is therefore due purely to self-energy effects.}, where only modest enhancement of $T_c$ is found near VHSs away from weak coupling \cite{radtke1993, radtke1994}. At sufficiently strong electron-phonon coupling, superconductivity is suppressed altogether by polaronic effects \cite{alexandrov2001,esterlis2018,esterlis2019,nosarzewski2021,chubukov2020}. Likewise, DQMC studies of superconductivity near an antiferromagnetic quantum critical point found that $T_c$ is insensitive to the presence of a VHS \cite{wang2017}. 

The enhancement of superconductivity by a VHS thus appears to be essentially a weak-coupling phenomenon, with interactions rapidly limiting its effectiveness. Consequently, strategies for achieving higher-$T_c$ in strongly correlated systems should not rely solely on engineering DOS singularities, but must account for the interplay between interactions and band structure. A maximum in $T_c$ at a given density is therefore not sufficient evidence for a VHS mechanism; concurrent observation of a sharp VHS in the DOS is required to establish this connection. 

Nevertheless, there are experimental systems in which interactions may not be weak, yet significant enhancements of $T_c$ are believed to arise from proximity to VHSs. Notable examples include Sr$_2$RuO$_4$ \cite{steppke2017, sunko2019, Mueller24}, kagome superconductors \cite{HuThomale22, Luo23}, and monolayer transition metal dichalcogenides \cite{WanUgeda23}. A key question is whether specific features of these systems enhance their sensitivity to VHSs, or whether they are effectively ``marginal," lying just on the weak-coupling side of the weak-to-strong coupling crossover. Alternatively, the enhancement of $T_c$ in some of these systems may arise from mechanisms unrelated to VHSs.

\begin{acknowledgments}
The authors thank Andrey Chubukov, Elio König, Carlos Sá de Melo, Joerg Schmalian, and Thomas Schäfer for helpful discussions. We also thank Dmitry Chichinadze and Steve Kivelson for useful comments on the manuscript.
This research was supported by the National Science Foundation (NSF) through the University of Wisconsin Materials Research Science and Engineering Center Grant No. DMR-2309000 (I. E. and G. R.). 
The work of D. S. and A. L. was supported by the NSF Quantum Leap Challenge Institute for Hybrid Quantum Architectures and Networks (Grant No. OMA-2016136) and by the H. I. Romnes Faculty Fellowship, provided by the University of Wisconsin–Madison Office of the Vice Chancellor for Research and Graduate Education, with funding from the Wisconsin Alumni Research Foundation.

\end{acknowledgments}

\appendix

\section{Analytic approximations}
\label{app:analytics}

In this appendix, we present a more detailed comparison of the DQMC results with analytic approximations. In addition to the perturbative results already presented in the main text, we also compare with the (unrenormalized) BCS approximation, where neither the Green's functions nor the vertex are dressed, and the RPA approximation, in which the dressed Green's function is used in the resummation of the ladder series for the pair susceptibility (see also Sec.~\ref{method:pert_theory}). 

\begin{figure}[h!]
    \centering
    \includegraphics[width=0.9\linewidth]{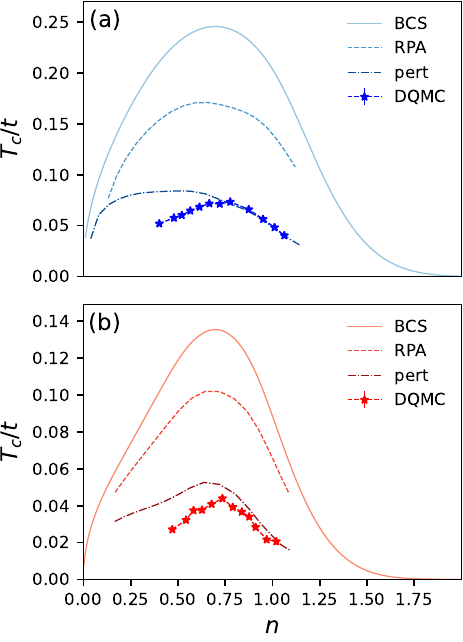}
    \caption{(a) Comparison of the DQMC $T_c$ with simple BCS, RPA, and second-order perturbation theory, for the ordinary VHS. Panel (a) is $U=-2t$ and panel (b) is $U = -1.5t$.}
    \label{fig:Tc_pert_comparison_oVH}
\end{figure}

\begin{figure}[h!]
    \centering
    \includegraphics[width=0.9\linewidth]{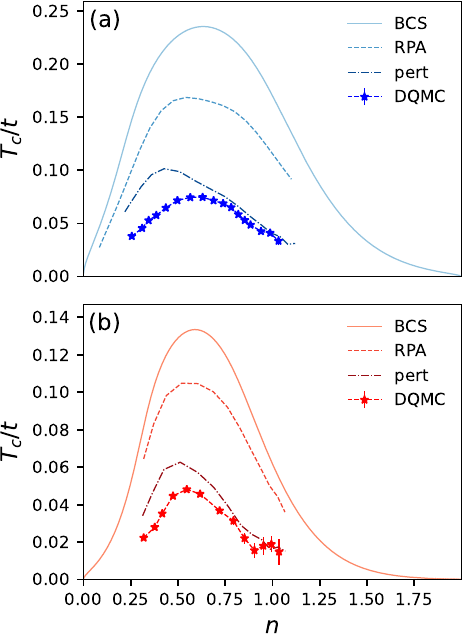}
    \caption{(a) Comparison of the DQMC $T_c$ with simple BCS, RPA, and second-order perturbation theory, for the HOHVS. Panel (a) is $U=-2t$ and panel (b) is $U = -1.5t$.}
    \label{fig:Tc_pert_comparison_hVH}
\end{figure}

Comparisons are shown for $U=-1.5t$ and $U=-2t$ in Figs.~\ref{fig:Tc_pert_comparison_oVH} and \ref{fig:Tc_pert_comparison_hVH}, for the case of the ordinary VHS and HOVHS, respectively. Unsurprisingly, the simplest BCS approximation grossly overestimates $T_c$. The RPA approximation, which takes into account the finite lifetime acquired by the electrons through the self-energy, decreases $T_c$, but still gives a significant overestimate. The addition of vertex corrections further suppresses the $T_c$. For $U=-1.5t$ the combination of self-energy and vertex corrections yields satisfactory qualitative agreement with DQMC over the density range considered, both for the ordinary VHS and the HOVHS. For $U=-2t$, the approximation is good for densities near and above the VHS density, but is of lower quality at densities below the VHS. In particular, the maximum $T_c$ is shifted to lower densities and, for the ordinary VHS, is excessively broadened. It is possible that these artifacts may be removed by extending the perturbative calculations to higher order in the interaction strength, but such calculations are beyond the scope of this work.

\section{Absence of competing density-wave instabilities}
\label{app:CDW}

\begin{figure}[h!]
    \centering
    \includegraphics[width=\linewidth]{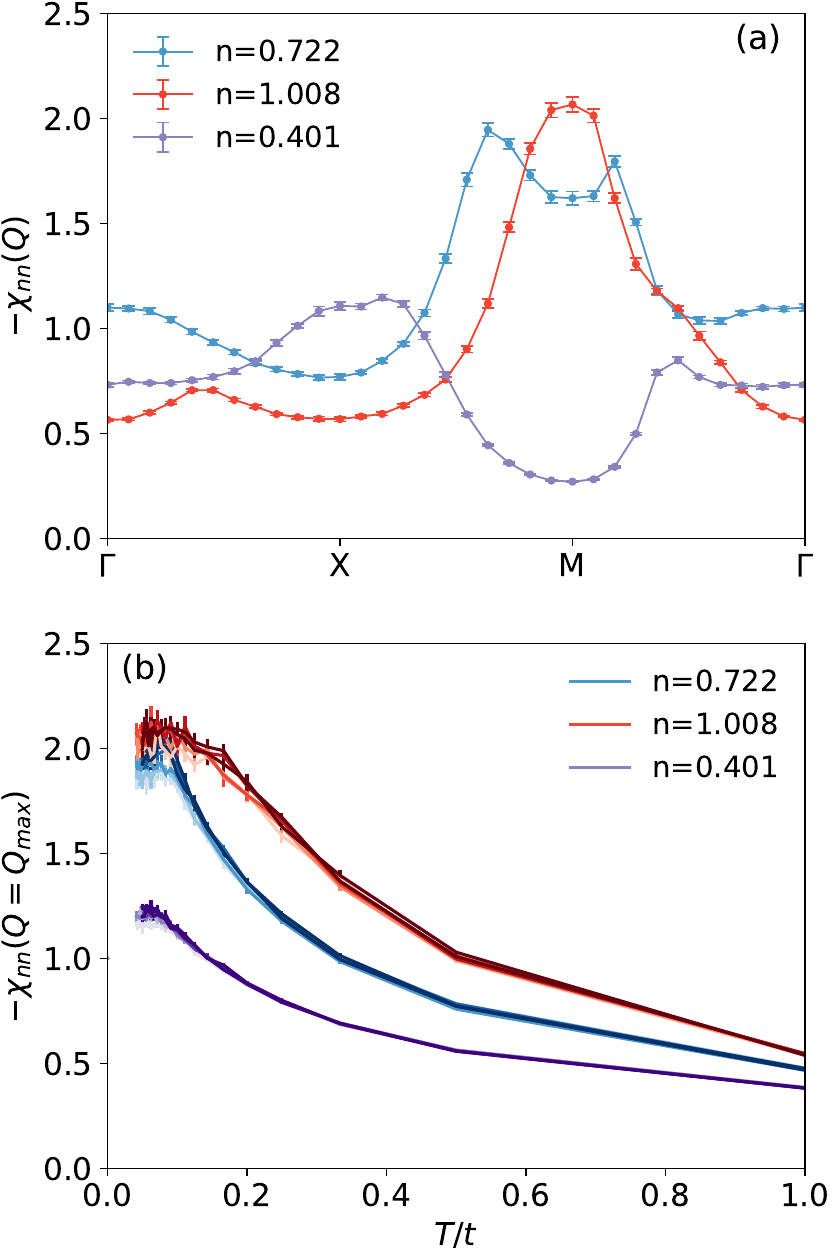}
    \caption{(a) Equal time charge susceptibility over a line-cut in the Brillouin zone for VHS filling (blue), half-filling (red) and a filling below VHS (purple) for parameters $|U|=2t,\, \beta t= 10 < \beta_ct,\, L=22$. (b) Static charge susceptibility as a function of $T/t$ for varying system size $L$ at momentum $\mathbf Q_{\text{max}}$ which maximizes the susceptibility for different fillings at $|U|=2t$. For each filling we have included data for linear system sizes $L=12 - 22$. We see that the peak is independent of the larger system sizes, indicating an absence of a divergence in the thermodynamic limit and we conclude an absence of charge order.}
    \label{fig:charge_susceptibility}
\end{figure}
In this appendix, we verify that the suppression of $T_c$ we observe is not due to incipient CDW order. This should be checked since,  for $t'=t''=0$, superconducting $T_c$ vanishes at the VHS due to the degeneracy between superconductivity and ${\mathbf Q}=(\pi,\pi)$ CDW order.
We probe the density correlations through the static charge susceptibility,
\be
    \chi_{nn}(\mathbf Q) = \frac 1N\sum_{\mathbf r}\int_0^\beta \dd\tau ~
    e^{-i\mathbf Q \cdot \bfr}
    \langle n(\mathbf r, \tau)n(0)\rangle,
\ee
where $n(\mathbf r) = c^\dagger_{\mathbf r\uparrow}c^{}_{\mathbf r\uparrow} + c^\dagger_{\mathbf r\downarrow}c^{}_{\mathbf r\downarrow}$.

We focus here on the case of the ordinary VHS ($t'=-0.3t$, $t''=0$). In  Fig.~\ref{fig:charge_susceptibility}a we report $\chi_{nn}(\mathbf Q)$ as function of $\mathbf Q$ for a few representative densities, including the VHS density $n\approx 0.72$. For $n\approx 1$, the charge susceptibility is indeed peaked near $\mathbf Q = (\pi,\pi)$, while the maximum shifts to other wave vectors with varying density. In Fig.~\ref{fig:charge_susceptibility}b we show the system-size and $T$-dependence of the charge susceptibility at the wave vector where it is maximal for a given density. It is clear that the charge susceptibility does not show any indication of divergence with increasing system size or lowering $T$, indicating that CDW order is not present in the system.

\section{Estimating $T_c$ from superfluid stiffness and pair correlations}
\label{app:extrap}

\begin{figure}[h!]
    \centering
    \includegraphics[width=\linewidth]{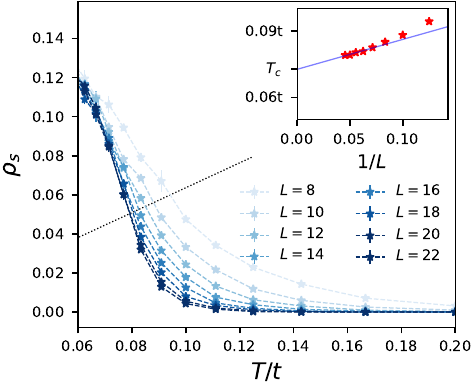}
    \caption{Superfluid stiffness $\rho_s$ as a function of $T$ for system sizes $L=8-22$. Here $U=-2t$ and the density is tuned to the ordinary VHS. The dashed line is $2T/\pi$. Inset: Red markers show the finite-size $T_c$ estimates from solving Eq.~\ref{eq:BKT}, together with the linear extrapolation to the $L\to\infty$ limit.}
    \label{fig:rho_s}
\end{figure}
In the main text, we have used the BKT criterion \eqref{eq:BKT}, based on the superfluid stiffness $\rho_s$,
 to extract $T_c$. In Fig.~\ref{fig:rho_s}
 we show our data for $\rho_s$ with varying system size, for the representative value $U=-2t$ at the VHS density. For each $L$ we obtain the finite-size estimate $T_c(L)$ by solving $\rho_s(T_c(L)) =2T_c(L)/\pi$. The inset of Fig.~\ref{fig:rho_s} shows the linear extrapolation of $T_c(L)$ to the thermodynamic limit $L\to\infty$. As noted in the main text, the extrapolation can be done more rigorously using  known logarithmic finite-size corrections to $\rho_s$ \cite{hsieh2013}, but we have found that extrapolation procedure to be less numerically stable, and we thus use a simple linear fit. Where the more rigorous extrapolation can be carried out reliably, we find it produces a result for $T_c$ indistinguishable from the linear extrapolation.

\begin{figure}[h!]
    \centering
    \includegraphics[width=\linewidth]{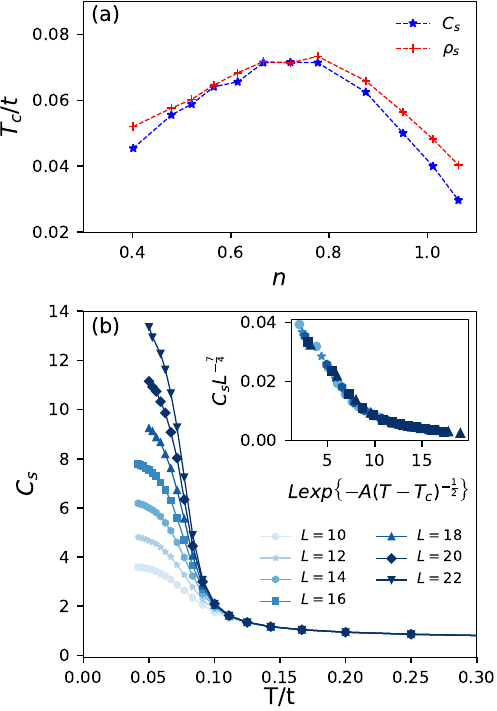}
    \caption{(a) Comparison of $T_c$ obtained from the BKT criterion and finite-size scaling of the pair correlation function $C_s$. (b) $C_s$ as a function of $T$ at the VHS density for system sizes $L=10-22$. The inset shows collapsed data using Eq.~\ref{eq:app_scaling} to obtain optimal values for $A$ and $T_c$.}
    \label{fig:scaling_Tc}
\end{figure}

 A complementary approach to extract $T_c$ is based on the finite-size scaling behavior of the equal-time pair correlation function,
\be
    C_s = \frac1N \sum_{\bfr} \langle \Delta^\dagger (\bfr)\Delta(0)\rangle, \quad \Delta(\bfr) = c_{\bfr \downarrow}c_{\bfr \uparrow},
\ee
which obeys the following finite-size-scaling law at a BKT transition \cite{moreoscalapino_scaling,paiva2004}:
\begin{equation}\label{eq:app_scaling}
    C_s = L^{\frac74}f(L/\xi), \quad \xi \sim \exp\left(-\frac{A}{\sqrt{T-T_c}}\right). 
\end{equation}
In this procedure, $T_c$ and $A$ are viewed as fitting parameters such that $C_s$ for varying $L$ collapses onto a universal curve. To determine these parameters systematically, we fit a continuous trial functions against the collapsed data and returned the $\chi^2$-value. The results of this procedure are shown in Fig.~\ref{fig:scaling_Tc}b,c. From Fig.~\ref{fig:scaling_Tc}a it can be seen that the estimate for $T_c$ based on the pair correlation function is in good agreement with that obtained from the BKT criterion on $\rho_s$.  

\section{Finite-size effects} \label{app: finite-size-effects-high-U} 
\begin{figure}[h!]
    \centering
    \includegraphics[width=\linewidth]{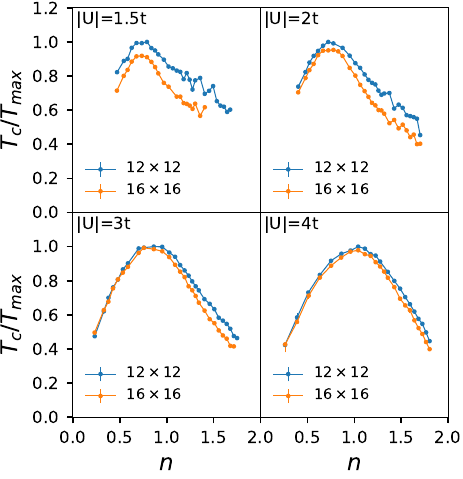}
    \caption{Normalized critical temperatures as a function of density for increasing values of the coupling, demonstrating the diminishing size effects with increasing $|U|$.}
    \label{fig:app-finite-size-Us}
\end{figure}

In the non-interacting limit $U\to 0$, the electronic eigenstates are plane waves and completely de-localized through the system. In the regime of small $U$, one thus expects finite-size effects to play an important role when interpreting numerical results. The opposite extreme of $|U|\to \infty$ corresponds to a localized single site problem and size effects are absent entirely. From this we expect that size effects discussed in Appendix \ref{app:extrap} will be less important as one increases the magnitude of $U$. This expectation is verified in Fig.~\ref{fig:app-finite-size-Us}, where we can see noticeable system size dependence of $T_c$ for $U = -1.5t$ and $U = -2t$, and much weaker size dependence when $|U| \geq 3t$. Thus, for $U = -1.5t$ and $U=-2t$, we have performed an extrapolation to the thermodynamic limit as described in Appendix~\ref{app:extrap}. For larger $|U|$, we simply use the finite size $T_c$ for the largest system size considered. 

We have attempted to further reduce size effects by threading a single magnetic flux quantum through the system, which destroys symmetries of the cluster and spreads out the non-interacting energy levels that otherwise would be frequently degenerate \cite{assaad2002}. The smaller energy level spacings yield smoother data as function of temperature. 

We introduce the uniform magnetic field $\mathbf B = B\mathbf e_z $ such that the vector potential is of the form $\mathbf A(x,y) = -\alpha By\mathbf e_x + (1-\alpha)Bx\mathbf e_y$ with $\alpha \in \mathbb R$ fixing the gauge. Performing a Peierls' substitution we can write the non-interacting part of the Hamiltonian as
\begin{equation}
    H = \sum_{ij}t_{ij}c^{\dagger}_ic_j^{}\exp\left\{iq\int_{(j_x,j_y)}^{(i_x,i_y)}d\mathbf r \cdot \mathbf A(\mathbf r)\right\},
\end{equation}
where $q$ is the charge of the electron. 
Periodic boundary conditions imply that the Hamiltonian should be invariant under translations $(i_x,i_y) \to (i_x+L_x, i_y)$ and $(i_x,i_y) \to (i_x, i_y+L_y)$.
By enforcing such periodicity we find the condition $qBL_xL_y = 2\pi n$ for $n\in \mathbb Z$, i.e., the lattice is threaded by an integer number of flux quanta. For the attractive Hubbard model, we thread a single flux quantum of opposite sign for spin-up and spin-down electrons, in order to preserve the anti-unitary symmetry that protects the fermion sign in DQMC \cite{li2019}.

\section{Equivalence between uniform and transverse electromagnetic response on the torus}\label{app: Static Uniform limit}
When extracting the superfluid stiffness $\rho_s$ from the DQMC data, as in Eq.~\ref{eq:rho_s}, we require the long wavelength limit of the static correlation function. Numerically this is subtle as the finite size of simulations imply a finite grid in $\mathbf q$-space. Typically, in DQMC literature, the smallest non-zero transverse $\mathbf q = (0, 2\pi/L)$ is used \cite{scalapino1993,paiva2004}.

However, on the torus, it turns out that setting $\mathbf q = (0, 0)$ gives the correct limit, i.e., the uniform and (limiting) transverse responses are the same for the superfluid stiffness. This equivalence is exploited to measure the helicity modulus in classical XY model simulations such as \cite{hsieh2013} but, to the best of our knowledge, has not been systematically proven.

To demonstrate the equivalence, we start with the Helmholtz-Hodge decomposition of the vector potential into three parts (longitudinal, transverse, harmonic/uniform):
\begin{equation}
    \mathbf A = \mathbf A_L +\mathbf A_T+\mathbf A_H,
\end{equation}
with $\nabla\cdot \mathbf A_T =0$, $\nabla \times \mathbf A_L = 0$, and $\mathbf A_H$ a constant. Such a decomposition is always possible on the torus, with the three components all orthogonal to each other (orthogonality defined with the real-space integral). Starting from the Ginzburg-Landau free energy functional with complex order parameter $\psi=\Delta e^{i\theta}$ and $\Delta$ a positive real constant, we have, in the presence of a vector potential,
\begin{equation}
    F[\theta, \mathbf A] = F_0 + \frac{\rho_s}{2}\int d^2\mathbf r \left(
    \nabla\theta-Q\mathbf A\right)^2.
\end{equation}
Here $Q = 2e$ is the charge of the condensate. Minimizing with respect to the phase $\theta$, we find
\begin{equation}
    \nabla^2\theta = Q\nabla\cdot\mathbf A.
\end{equation}
By going to momentum space $\nabla \to i\mathbf k$ we find that this implies
\begin{equation}
    \nabla \theta \to i\mathbf k\theta  = Q \hat{\mathbf k} (\hat{\mathbf k}\cdot\mathbf A) \equiv Q\mathbf A_L.
\end{equation}
Here the hat indicates a normalized vector. This highlights that the $\mathbf A_L$ component is pure gauge. In contrast, $\mathbf A_H$ cannot be gauged away on the torus: if one tries to write $\mathbf A_H = \nabla\left( \mathbf A_H \cdot \mathbf r\right)$, the scalar $\mathbf A_H \cdot \mathbf r$ does not respect the torus periodicity.

Inserting $\nabla \theta$ back into the free energy, we find, as a functional of the vector potential,
\begin{align*}
        F_{\text{min}}[\mathbf A] &= F_0 + \frac{\rho_s Q^2}{2}\int d^2\mathbf r \left(\mathbf A_T + \mathbf A_H\right)^2 \\
        &= F_0 + \frac{\rho_s Q^2}{2}\int d^2\mathbf r \left(\mathbf A_T^2 + \mathbf A_H^2\right),
\end{align*}
where we used orthogonality to neglect the cross-term. Since the coefficient is the same for $\mathbf A_H$ and $\mathbf A_T$, we conclude that the response to a homogeneous vector potential is the same as the response to a transverse vector potential.

\section{Strong-coupling self-energy}
\label{app: strong coupling self-energy}

\begin{figure}[]
    \centering
    \includegraphics[width=1.0\linewidth]{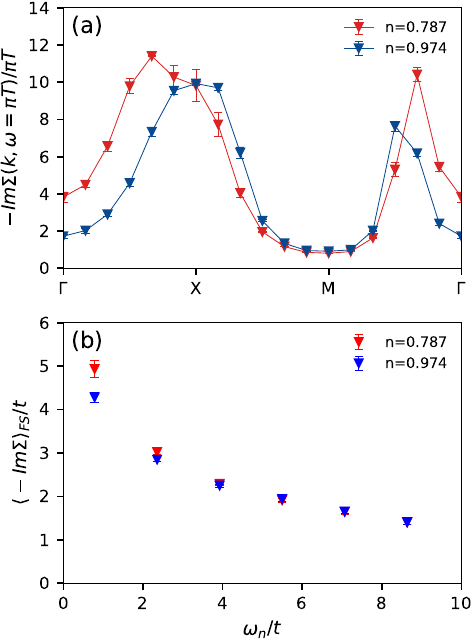}
    \caption{(a) Imaginary part of the self-energy for $|U|=8t$ at lowest Matsubara frequency for a density near the VHS ($n\approx 0.79$) and for a density away from the VHS ($n \approx 0.97$). We that the self-energy is not too dependent on the filling however there is still momentum-space structure. (b) Momentum average of the imaginary part of the self energy as a function of Matsubara frequency for the same fillings as (a).}
    \label{fig:app self-energy-strong-coupling}
\end{figure}

In Figs.~\ref{fig:sig} and \ref{fig:sig_hvh} of the main text we have reported the electronic self-energy for the representative weak-coupling $U=-1.5t$ with Fermi-surface average defined as,
\begin{align}
    \langle B\rangle_{\text{FS}} &= \frac{1}{\sum_\mathbf k \Delta_\beta (\mathbf k)}\sum_\mathbf k \Delta_\beta (\mathbf k) B(\mathbf k),\\
    \Delta_\beta (\mathbf k)&= -\frac{1}{\beta}\frac{\partial f(\epsilon_\mathbf k)}{\partial\epsilon_\mathbf k},
\end{align}
where $f(\epsilon) = \left[1+e^{\beta(\epsilon-\mu)}\right]^{-1}$ is the usual Fermi function. 

For completeness, in this appendix we also show the self-energy at stronger coupling, Here we choose a uniform momentum averaging procedure ($\Delta_\beta (\mathbf k) = 1$), since we do not expect the non-interacting Fermi surface to play a significant role in the strong-coupling regime. 

In Fig.~\ref{fig:app self-energy-strong-coupling}, we show the imaginary part of the self-energy for $U=-8t$. Panel (a) displays the $k$-dependence at the lowest Matsubara frequency. The pronounced momentum-space structure at this relatively large $|U|$ is somewhat unexpected and may be related to the redistribution of spectral weight from the opening of the strong-coupling pseudogap. Panel (b) shows the momentum-averaged Matsubara-frequency dependence, which increases monotonically with decreasing frequency. This behavior is consistent with strong-coupling expectations; in the atomic limit, $\Sigma(i\omega_n) \sim U^2/i\omega_n$.

\section{Proxies for the interacting density of states}
\label{app:DOS_proxies}

\begin{figure}
    \centering
    \includegraphics[width=1.0\linewidth]{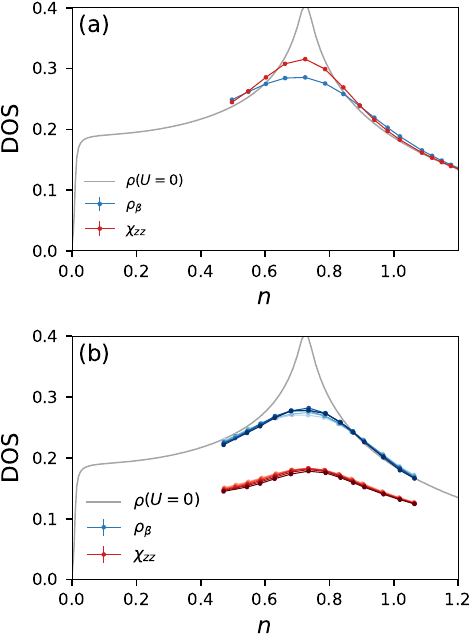}
    \caption{(a) The DOS proxy $\rho_\beta$ defined in Eq.~\eqref{eq:DOS_proxy} and the spin susceptibility $\chi_{zz}$, together with the non-interacting exact DOS (solid line) for $|U|=0t$. The data is shown for temperature $\beta t = 10$. (b) Same as (a) but for $|U|=1.5t$ and varying temperatures $\beta t = 11 -16$ above the critical temperature. Interactions renormalize the spin susceptibility much more drastically than $\rho_\beta$ but they both agree on the location of the VHS.} 
    \label{fig:app_DOS_proxies}
\end{figure}

\begin{figure}[htbp]
    \centering
    \includegraphics[width=1.0\linewidth]{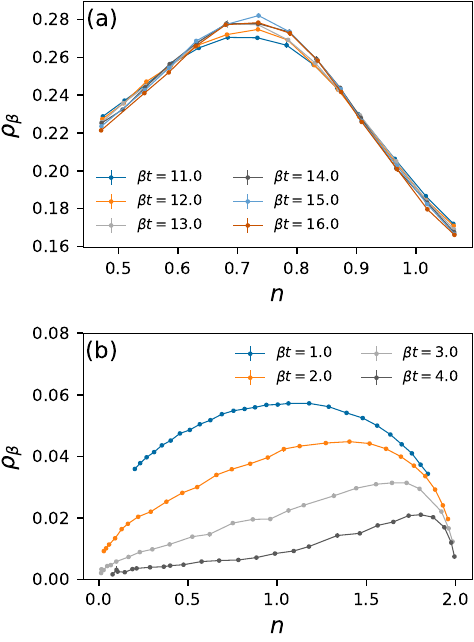}
    \caption{Comparison of the DOS proxy $\rho_\beta$ for (a) weak coupling, $U=-1.5t$ and (b) strong coupling, $U=-8t$. At stronger coupling, $\rho_\beta$ is still strongly temperature dependent near the critical temperature. In both panels, all temperatures are above the superconducting critical temperature ($T_c \approx 0.06t$ for $U=-1.5t$ and $T_c\approx 0.20t$ for $U=-8t$).}
    \label{fig:app_strong_weak}
\end{figure}

The location of the VHS density for $U\neq 0$ was identified using the imaginary-time observable $\rho_\beta$, defined in  Eq.~\eqref{eq:DOS_proxy}. Since $\rho_\beta$ is not a direct measure of the electronic DOS but rather a proxy for it, we have also measured another common proxy for the DOS in a Fermi-liquid---the low-temperature spin susceptibility:
\be
    \chi_{zz} = \frac 1N \sum_{\mathbf r}\int_0^\beta d\tau \langle S^{z}( \mathbf r, \tau)S^{z}(0)\rangle, 
\ee
where $S^z(\bfr) = \frac 12 (n_{\bfr \uparrow} - n_{\bfr \downarrow})$. The different proxies are compared in Fig.~\ref{fig:app_DOS_proxies}a for $U=0$ and in Fig.~\ref{fig:app_DOS_proxies}b for $U=-1.5t$. While the overall magnitudes of $\rho_\beta$ and $\chi_{zz}$ differ for $U\neq 0$, they both exhibit a well-defined maximum near the same density, which coincides with the VHS density of the non-interacting system. 

In Fig.~\ref{fig:app_strong_weak} we compare $\rho_\beta$ at weak and strong coupling. At weak coupling, its low-temperature value provides a meaningful estimate of the electronic DOS, whereas at strong-coupling $\rho_\beta$ acquires a pronounced temperature dependence and exhibits structure unrelated to the non-interacting DOS. In the latter regime, the evolution of $\rho_\beta$ is presumably tied to the emergence of spectral features associated with the onset of a pseudogap.

\section{Strong-coupling expansion}
\label{app: Strong coupling expansion}

\begin{figure*}
    \centering
    \includegraphics[width=\textwidth]{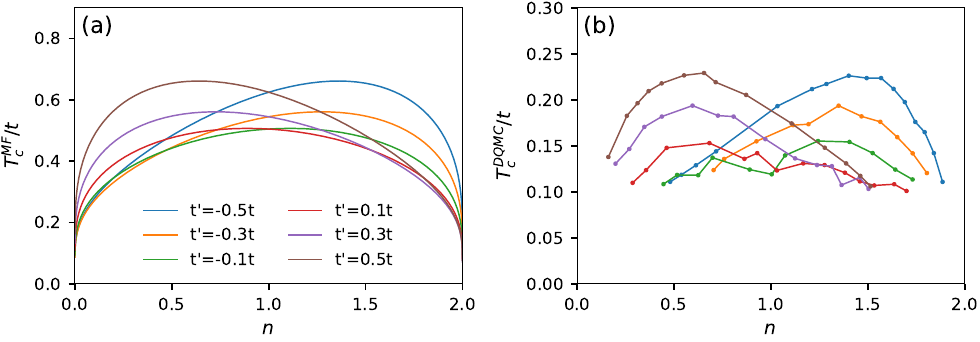}
    \caption{(a) The mean-field $T_c$, Eq.~\eqref{eq:Tc_MF}, for $U=-8t$ and for varying values of $t'/t$. (b) DQMC results for the same parameters. Note the strong $t'$ dependence of the optimal density and overall shape of $T_c(n)$, which is captured by the strong-coupling, mean-field prediction.}
    \label{fig:strong coupling Tc varying tp}
\end{figure*}

In this appendix, we provide details of the strong-coupling expansion of the attractive Hubbard model, following closely Ref.~\cite{Freericks1998}.

When $|U|\to\infty$, all electrons are paired up, with each site either doubly occupied or empty. The ground state energy will not depend on where the electron pairs are positioned on the lattice, leading to a large degeneracy. The degeneracy is lifted by the kinetic energy term. We write 
\begin{equation}
    H = H_0 + T,
\end{equation}
with $H_0 = -U\sum_i n_{i\uparrow}n_{i\downarrow}$ being the unperturbed Hamiltonian and $T = -\sum_{ij\sigma}t_{ij}c^\dagger_{i\sigma}c^{}_{j\sigma}$ being the perturbation. 

We make use of the projector approach \cite{Freericks1998}. Define projection operators $P_0$ and $P$ as projectors onto the ground state of $H_0$ and $H$, respectively, and define the overlap operator $O=P_0PP_0$. The effective Hamiltonian can be constructed in powers of $T$ as
\begin{equation}
    H_{\text{eff}} = O^{-1/2}P_0HPP_0O^{-1/2},
\end{equation}
with
\begin{align}
    O^{-1/2} &= P_0 +\sum_{n=1}^\infty \frac{(2n-1)!!}{(2n)!!}[P_0 - O]^n, \\
    P &= P_0 -\sum_{n=1}^\infty\sum_{k_i\ge0}{}{}^{'}R^{k_1}TR^{k_2}T\dots TR^{k_{n+1}},
\end{align}
the primed sum indicating summation only over terms satisfying $\sum_ik_i=n$ with integer $k_i$. The operator $R$ is defined as
\begin{equation}
    R^0\equiv-P_0, \quad R^k = \left(\frac{1-P_0}{E_0-H_0}\right)^k, \quad  (k>0).
\end{equation}
Making use of $P_0R = RP_0 = 0 = P_0TP_0$ we find, to third order in $T$,
\begin{widetext}
\begin{align}
H_{\text{eff}} &= -(2\mu+|U|)\sum_ib^
\dagger_ib^{}_i -\frac{2t^2}{|U|}\sum_{\langle ij\rangle}\left[b^\dagger_ib_j +n_j(1-n_i) + \text{h.c.}\right] -\frac{2t'^2}{|U|}\sum_{\langle\langle ij\rangle\rangle}\left[ b^\dagger_ib_j +n_j(1-n_i) +\text{h.c.}\right] \\
&\qquad -\frac{2t^2t'}{U^2}\sum_{\langle\langle ij\rangle\rangle}\sum_{\langle k, ij\rangle}\left[3b^\dagger_ib_j+2n_j(1-n_i) +\text{h.c.}\right] \left(1-n_k\right).
\end{align}
\end{widetext}
Here $b_i = c_{i\downarrow}c_{i\uparrow}$ annihilates a hard-core boson at site $i$ and $n_i = b^\dagger_ib^{}_i$ is the boson occupation number. In the last term, the sum is over all $k$ that are simultaneous nearest neighbors to $i$ and $j$. 

For a mean-field analysis, it is useful to express $H_\text{eff}$ in pseudo-spin variables by introducing $J_i^+ = b^\dagger_i$, $J_i^-=b_i$, and $J_i^z = n_i-1/2$. In these variables, the Hamiltonian can be written
\begin{widetext}
    \begin{align}
H_{\text{eff}} &= -(2\mu+|U|)\sum_iJ_i^z +\frac{4t^2}{|U|}\sum_{\langle ij\rangle}\left\{2J_i^zJ_j^z - \mathbf J_i\cdot \mathbf J_j -\frac14\right\} 
+\frac{4t'^2}{|U|}\sum_{\langle\langle ij\rangle\rangle}\left\{2J_i^zJ_j^z - \mathbf J_i\cdot \mathbf J_j -\frac14\right\} \\
&\qquad -\frac{4t^2t'}{U^2}\sum_{\langle\langle ij\rangle\rangle}\sum_{\langle k, ij\rangle}\left(3\mathbf J_i\cdot \mathbf J_j - 4J_i^zJ_j^z+\frac12\right)\left(\frac12-J^z_k\right).
\end{align}
\end{widetext}

With this Hamiltonian we are ready to perform the standard mean-field procedure. Writing $J_i = \langle J_i\rangle + \delta J_i$ with $\delta J_i = J_i - \langle J_i \rangle$ and expanding the effective Hamiltonian to linear order in $\delta J_i$, we find 
\begin{align}
    H_{\text{MF}} &= -\sum_i \mathbf h \cdot \mathbf J_i, \\
    h_x &= \frac{16}{|U|}\left[t^2+t'^2 + \frac{6t^2t'}{|U|}\left(\frac12-m_z\right)\right]m_x \equiv Am_x,
\end{align}
where $m_{x} = \langle J^{x}\rangle$ is a mean-field corresponding to superfluid XY-order in the boson description \cite{micnas1990,Freericks1998}.
The self-consistency equation becomes
\begin{equation}
    \mathbf m = \frac{1}{2}\frac{\mathbf h}{|\mathbf h|}\tanh\left(\frac{\beta}{2}|\mathbf h|\right).
\end{equation}
Linearizing this equation near $T=T_c$, we obtain
\begin{align}
    h_z &= \frac A2\tanh\left(\frac{\beta_c}{2}h_z\right), \\
    m_z &= \frac12 \tanh\left(\frac{\beta_c}{2}h_z\right).
\end{align}
Solving for $T_c$ and setting $2m_z=n-1 $, we obtain
\begin{equation}
    T_c = \frac{(n-1)}{\ln\left(\frac{n}{2-n}\right)}\frac{8}{|U|}\left[t^2+t'^2 +\frac{6t^2t'}{|U|}(1-n)\right],
\end{equation}
which is Eq.~\eqref{eq:Tc_MF} of the main text. 
In Fig.~\ref{fig:strong coupling Tc varying tp}, the mean-field $T_c$ is compared to DQMC data for varying values of $t'$ at strong coupling $U=-8t$. The figure illustrates that the location of the maximum in $T_c(n)$ for different $t'$ tracks the strong-coupling prediction. 
\bibliography{van_Hoves}

\end{document}